\documentclass[12pt]{article}
\input{quantum-d-sep.sty}

\begin{document}
\title{Quantum d-separation\\
and\\
quantum belief propagation}
\date{ \today}
\author{Robert R. Tucci\\
        tucci@ar-tiste.com}
\maketitle
\vskip2cm
\section*{Abstract}
The goal of this paper is to generalize
classical d-separation
 and 
classical Belief Propagation (BP)
 to the quantum realm.
Classical d-separation
is 
an essential ingredient of most of Judea Pearl's 
work. It 
is crucial
to all 3 rungs of 
what Pearl calls the 3 rungs
of Causation.
So having a quantum version
of d-separation
and BP
probably 
implies
that most of Pearl's
Bayesian networks
work, 
including
his theory of
causality, 
can be translated
in a straightforward 
manner to
the quantum realm.
\newpage
\section{Introduction}

Classical Bayesian networks (bnets),
 d-separation
and belief propagation (BP), the main topics 
of this paper, were first proposed  
by Judea Pearl and collaborators. Pearl's results 
and their history
have been amply discussed by
him and some coauthors in 
the following 4 books:
Refs.\cite{pearl-1988book,pearl-2013book,
pearl-primer,book-why}.
Tucci has written 
 an open source book entitled
Bayesuvius (see Ref.\cite{bayesuvius}) about
classical bnets.
Bayesuvius will
be cited frequently in this paper as
a primary source of 
background information
explained in the same style
of notation as this paper. 

Quantum Bayesian Networks (qbnets) were 
first proposed by Tucci in Ref.\cite{qbnets1}.
Ref.\cite{qbnets1} deals with pure quantum states only.
Tucci later generalized qbnets 
to mixed states
in
Ref.\cite{qbnets-mixed}.

The goal of this paper is to generalize
classical d-separation\footnote{Discussed 
in chapter of Bayesuvius entitled 
``D-Separation".}
 and 
classical BP\footnote{Discussed 
in chapter of Bayesuvius entitled 
``Message Passing (Belief Propagation)".}
 to the quantum realm.
Tucci has assumed or implied
in previous work that d-separation
and BP
are valid for qbnets, but he
has never explicitly proven  this.
This paper is intended to be a first
step towards filling that gap.

Classical d-separation inspired
Tucci to propose the definition
of squashed entanglement (SE).
d-separation and SE
are very closely linked. 
SE
was first proposed  by Tucci
in a series of 6 papers 
Refs.\cite{ent1, ent2, ent3, ent4, ent5, ent6}
spanning the years 1999-2002.
Starting around 2004 with
Ref.\cite{putrid-squash} by Christiandl and Winters,
other researchers 
recognized the importance of SE
and began writing papers about it.
A more complete history of SE
is given in Ref.\cite{wiki-sq-ent}.
Subsequently, Tucci has written an open source software
program 
called Entanglish (see Ref.\cite{entanglish}) for 
calculating SE.

The open-source computer program
 Quantum Fog (Ref.\cite{qfog})
written by Tucci assumes
that
both classical and quantum BP
 are valid, because
it uses the 
junction tree (JT) algorithm  
to do both classical
and quantum inference with bnets.
The JT
algorithm\footnote{Discussed 
in chapter of Bayesuvius entitled 
``Juntion Tree Algorithm".}
is a generalization of BP so as to 
include bnets with loops.\footnote{
The quantum version of JT used by the current version of
Quantum Fog is not quite correct
because it does not use the vector amplitudes defined
in this paper.}

Classical d-separation
is 
an essential ingredient of most of Judea Pearl's 
work. It 
is crucial
to all 3 rungs of 
what Pearl calls the 3 rungs
of Causation.
So having a quantum version
of d-separation
and BP
probably 
implies
that most of Pearl's
bnet
work, 
including
his theory of
causality, 
can be translated
in a straightforward 
manner to
the quantum realm.
This
is perhaps not too surprising
because most
of Pearl's bnet 
results depend to
a large extent
on the topology (i.e., graph
structure) of the underlying
DAG of a bnet, and that DAG 
should be essentially
the same whether we are
considering a quantum
phenomenon or
its classical limit.

\section{Notational Conventions
 and Preliminaries}
This paper
will employ the same
notational
conventions as
Bayeuvius (Ref.\cite{bayesuvius}).
Hence, if
the reader encounters
any notation that
is not defined
in this paper,
he/she should consult Bayesuvius (especially
its chapter
entitled ``Notational Conventions and Preliminaries"),
where it is very likely
to be defined.

Henceforth,
whenever
we write
$[\cale][h.c.]$,
where $\cale$
is some quantum operator,
we will mean $\cale\cale^\dag$,
where $\cale^\dag$
is the Hermitian conjugate
of $\cale$.
Also, whenever
we write
$\caln(!x)$,
we will
mean a 
normalization
constant that is independent
of $x$.

Consider
a DAG with $nx$ nodes
$\rvx^{nx}= (\rvx_i)_{i=0, 1, \ldots, nx-1}$.
A qbnet 
consists of such a DAG with 
a TPM (transition probability matrix)
attached to each node,
but unlike the TPMs for a classical bnet,
these ones are complex valued
and normalized differently.
If we represent the TPM
of node $\rvx_j$
by a {\bf probability 
amplitude} $A(x_j|pa(\rvx_j))$,
then $A(x_j|pa(\rvx_j))$ must satisfy
the normalization condition

\beq
\sum_{x_j}
\abs{ A(x_j|pa(\rvx_j))}^2=1
\;.
\eeq
The amplitude for the full
DAG is defined as in the classical
case, by multiplying
the TPMs of all the nodes:

\beq
A(x^{nx})
=
\prod_j A(x_j|pa(\rvx_j))
\;.
\eeq
Note that

\beq
\sum_{x^{nx}}\abs{A(x^{nx})}^2=1
\;.
\eeq

Suppose
$\rva., \rvb.$ are disjoint multinodes 
in $\rvx^{nx}$. Denote
the complement
multinode of $\rva.$ by
$\ul{a_.}^c=\rvx^{nx}-\rva.$ Then
we define
the {\bf vector amplitude} $\A(a.)$ 
and the probability $P(a.)$ by

\beq 
\A(a.)=\ket{A(a.)}=\ket{a.}\sum_{a^c_.}A(a., a.^c)\ket{a.^c}
\;,
\eeq
with boundary cases
$\A(x^{nx})=A(x^{nx})$,\;
$\A=\sum_{x^{nx}}A(x^{nx})\ket{x^{nx}}$.
and

\beq
P(a.)=
\sum_{a.^c}\abs{A(a., a.^c)}^2
\;.
\eeq
Note that $\sum_{a.} P(a.)=1$
so $P(a.)$ is a bona
fide probability
distribution as the notation implies. Note also that 

\beq
\av{A(a.)|A(a.)}= P(a.)
\;, 
\eeq
so you
can think of $\A(a.)$
as being a generalized square root of $P(a.)$.

We define the {\bf conditional vector amplitude} $\A(b.|a.)$ by
\beq
\A(b.|a.)=\ket{A(b.|a.)}=
\frac{\A(b., a.)}{\A(a.)}
\;.
\eeq
This is analogous to the definition
of conditional
probability,
$P(b.|a.)=
\frac{P(b., a.)}{P(a.)}$.
Note that $\A(b.|a.)$
isn't really a ket; it's a tuple
of two kets, but it is convenient
to represent it as a ket.
We can also define the dual $\bA(b.|a.)$
of $\A(b.|a.)$ so that
the following equations are satisfied:

\beq
\bra{A}(b.|a.)=\bra{A(b.|a.)}=
\frac{\bA(b., a.)}{\bA(a.)}
\;,
\eeq

\beq
\av{A(b.|a.)|A(b.|a.)}=
\frac{\av{A(b.,a.)|A(b., a.)}}
{\av{A(a.)|A(a.)}}=
\frac{P(b., a.)}{P(a.)}=
P(b.|a.)
\;.
\eeq
Hence, you can think of 
$\A(b.|a.)$
as being a generalized square root
of $P(b.|a.)$.

Suppose $\rva., \rvb., \rvc.$
are disjoint multinodes such that
$\rva.\cup\rvb.\cup\rvc.=\rvx^{nx}$ Then
\beqa
\sum_{b.} \A(a., b.)
&=&
\sum_{b.} \sum_{c.}\A(a., b., c.)
\\
&=&
\A(a.)
\eeqa
This is analogous to the probability  marginalization 
rule $\sum_{b.}P(a., b.)=P(a.)$.

Note that
\beqa
\sum_{b.} \A(a.|b.)\A(b.)
&=&
\sum_{b.} \A(a., b.)
\\
&=&
\A(a.)
\;.
\eeqa
This
is analogous to
the  probability splitting rule $P(a.)=\sum_{b.}P(a.|b.)P(b.)$.

Suppose $\rva.$ and $\rve.$ are
disjoint multinodes,
with $\rve.$
representing evidence. Note that
\beqa
\A(a.|e.)&=&
\frac{\A(a., e.)}{\A(e.)}
\\
&=&
\frac{\A(e.|a.)\A(a.)}{\A(e.)}
\;.
\eeqa
This is analogous
to the classical Bayes rule 
$P(a.|e.)=\caln(!a.)P(e.|a.)P(a.)$.

Let $\calp_\rvlam$ 
be the 
set of all 
probability 
distributions $P_\rvlam(\lam)$
with $\lam \in S_\rvlam$.

Let $\calh_\rvx$ represent 
a Hilbert space spanned
by an orthonormal basis $\ket{x}$,
where $x\in S_\rvx$.
Let $\calh_{\rvx, \rvy}=
\calh_\rvx\otimes\calh_\rvy$.

Let $\cald_\rvx$ be the set of all
 density matrices acting on $\calh_\rvx$.
Likewise, let $\cald_{\rvx, \rvy}$
be the set of all density matrices
acting on $\calh_{\rvx, \rvy}$.
If $\rho_{\rvx,\rvy}\in \cald_{\rvx,\rvy}$,
then $\rho_\rvx$
will denote the partial trace
$\tr_\rvy \rho_{\rvx, \rvy}$ .

Let $\cald_{\rvx, \rvy, \rvlam^d}$
be the set of all
density matrices in
$\cald_{\rvx, \rvy, \rvlam}$
which are diagonal in $\lam$.
In other words, $\rho_{\rvx,\rvy, \rvlam^d}
\in \cald_{\rvx, \rvy, \rvlam^d}$
if it is of the form

\beq
\rho_{\rvx,\rvy, \rvlam^d}=
\sum_\lam P_\rvlam(\lam)\ket{\lam}\bra{\lam}
\rho^\lam_{\rvx, \rvy}
\;,
\eeq
where
 $\rho^\lam_{\rvx, \rvy} \in \cald_{\rvx,\rvy}$
for all $\lam$
and $P_\rvlam\in \calp_\rvlam$.

\begin{table}[h!]
\centering
\begin{tabular}{|
>{\columncolor[HTML]{ECF4FF}}l |l|l|}
\hline
 & \cellcolor[HTML]{ECF4FF}classical & \cellcolor[HTML]{ECF4FF}quantum \\ \hline
Entropy & $H(\rvx)=-\sum_x P(x)\ln P(x)$ & \begin{tabular}[c]{@{}l@{}}$S_{\rho}(\rvx)=S(\rho)=-\tr_\rvx( \rho \ln \rho)$\\ for $\rho\in \cald_\rvx$\end{tabular} \\ \hline
\begin{tabular}[c]{@{}l@{}}Conditional\\ Entropy\end{tabular} & $H(\rvx|\rvy)=-\sum_{x,y}P(x,y)\ln P(x|y)$ & \begin{tabular}[c]{@{}l@{}}$S_{\rho}(\rvx|\rvy)=-S_\rho(\rvy)+S_\rho(\rvx,\rvy)$\\ for $\rho\in \cald_{\rvx,\rvy}$\end{tabular} \\ \hline
\begin{tabular}[c]{@{}l@{}}Mutual\\ Information\end{tabular} & $H(\rvx:\rvy)=\sum_{x,y}P(x,y)\ln \frac{P(x,y)}{P(x)P(y)}$ & \begin{tabular}[c]{@{}l@{}}$S_{\rho}(\rvx:\rvy)=S_\rho(\rvx)+S_\rho(\rvy)-S_\rho(\rvx,\rvy)$\\ for $\rho\in \cald_{\rvx,\rvy}$\end{tabular} \\ \hline
\begin{tabular}[c]{@{}l@{}}Conditional\\ Mutual\\ Information\end{tabular} & \begin{tabular}[c]{@{}l@{}}$H(\rvx:\rvy|\rvlam)=$\\ $\sum_{x,y,\lam}P(x,y,\lam)\ln \frac{P(x,y|\lam)}{P(x|\lam)P(y|\lam)}$\end{tabular} & \begin{tabular}[c]{@{}l@{}}$S_{\rho}(\rvx:\rvy|\lam)=S_\rho(\rvx|\lam)+S_\rho(\rvy|\rvlam)-S_\rho(\rvx,\rvy|\rvlam)$\\ for $\rho\in\cald_{\rvx,\rvy, \rvlam}$\end{tabular} \\ \hline
\end{tabular}
\caption{Definitions of various Entropies and Informations.}
\label{tab-quan-clas-info}
\end{table}
Table \ref{tab-quan-clas-info}
gives the 
definitions 
of various entropies and
informations
used in 
classical Shannon
Information Theory (SIT),
and their counterparts
in quantum SIT.

\section{Quantum d-separation}

The next box comes
from the chapter of Bayesuvius
entitled ``D-Separation".

\begin{framed}
\begin{claim}(Classical d-separation Theorem)

Suppose
$\rvA., \rvB., \rvZ.$
are disjoint multinodes
of a DAG  $G$.

$\rvA.\perp_G \rvB.|\rvZ.$ iff
$H(\rvA. : \rvB.|\rvZ.)=0$
for all $P$
compatible with $G$.

\end{claim}
\end{framed}
The proof
of this theorem
will
not be presented
here. It isn't
presented in the 
current version of Bayesuvius either.
To see it, 
you will
have to look
in Ref.\cite{pearl-1988book},
and in the original 
papers cited therein.

The definition
of classical d-separation 
(i.e., of $\rvA.\perp_G \rvB.|\rvZ.$)
only depends on the topology
of the DAG. We will
define quantum d-separation
exactly as it
is
defined classically.
Whether the bnet has 
probabilities
or amplitudes
for its TPMs
does not make a difference
at the 
level
of the definition
of d-separation.
It only becomes
important
when trying
to
find a quantum
analogue of the 
whole classical
d-separation {\it theorem}
which is stated in the box above.
The remainder
of this section
will
be dedicated to finding 
a quantum
analogue
to the whole box above.

We start by
using
the definitions
introduced
in the previous section
to conclude that:

\beq
S_\rho(\rvx:\rvy|\rvlam^d)=
\sum_\lam P(\lam) \left[
S(\rho^\lam_x)+S(\rho^\lam_y)-
S(\rho^\lam_{\rvx,\rvy})
\right]
\eeq
for $\rho\in \cald_{\rvx,\rvy, \rvlam^d}$.

\begin{claim}
$S_\rho(\rvx:\rvy|\rvlam^d)=0$
iff $\rho^\lam_{\rvx, \rvy}=
\rho^\lam_{\rvx}
\rho^\lam_{\rvy}$
for all $\lam$.
\end{claim}
\proof

$S(\rho^\lam_x)+S(\rho^\lam_y)-
S(\rho^\lam_{\rvx,\rvy})=0$
iff
$\rho^\lam_{\rvx, \rvy}=
\rho^\lam_{\rvx}
\rho^\lam_{\rvy}$
\qed

Next, we express 
each density matrix
$\rho^\lam_{\rvx, \rvy}$ as
\beq
\rho^\lam_{\rvx, \rvy}
=U D U^\dag
\;,
\eeq
where $U$
is a unitary
matrix
and $D$
is a diagonal matrix
with non-negative diagonal entries.
Now let

\beq
A(x,y|x_0, y_0, \lam)=
\bra{x}\bra{y}U\ket{x_0}\ket{y_0}
\eeq 
and

\beq
\bra{x_0}\bra{y_0}
D
\ket{x_0}\ket{y_0}
=
P(x_0, y_0|\lam)
\;.
\eeq
Hence,

\beq
\rho^\lam_{\rvx, \rvy}
=
\sum_{x_0,y_0} 
\Big[
\sum_{x,y}
\ket{x}\ket{y}
A(x,y|x_0, y_0, \lam) 
\Big]
P(x_0,y_0|\lam)
\Big[
h.c.
\Big]
\eeq

To make our
future
expressions
more concise, define 
the two abbreviations
$R_0=(x_0, y_0)$,
$R=(x, y)$. Then

\beq
\rho_{\rvx, \rvy, \rvlam^d}
=
\sum_{R_0,\lam} 
\Big[\ket{\lam}
\sum_{R}
\ket{R}
\underbrace{A(R|R_0, \lam)
\sqrt{P(R_0|\lam)}
\sqrt{P(\lam)}}_{A(R,R_0,\lam)}
\Big]
\Big[
h.c.
\Big]
\;.
\eeq
From the
definitions 
of conditional
probabilities
and conditional amplitudes,
we get

\beq
A(R|R_0, \lam)=
A(y|x, R_0, \lam)A(x|R_0, \lam)
\eeq
and

\beq
P(R_0|\lam)=
P(y_0|x_0,\lam)P(x_0|\lam)
\;.
\eeq
Therefore,

\beq
A(R,R_0,\lam)\;=\;
A(y|x, R_0, \lam)A(x|R_0, \lam)
\sqrt{P(y_0|x_0, \lam)P(x_0|\lam) P(\lam)}
\;.
\label{eq-full-a-sq-ent}
\eeq

\begin{figure}
$$
\begin{array}{ccc}
\xymatrix{
\rvy_0\ar[dd]\ar@/^1pc/[ddrr]
&&\rvx_0\ar[dd]\ar[ll]\ar@/_1pc/[ddll]
\\
&\rvlam\ar[dl]\ar[dr]
\ar[ul]\ar[ur]
\\
\rvy&&\rvx\ar[ll]
}
&\;\;\;&
\xymatrix{
\rvy_0\ar[dd]&&\rvx_0\ar[dd]
\\
&\rvlam\ar[dl]\ar[dr]
\ar[ul]\ar[ur]
\\
\rvy&&\rvx
}
\\
(a)
&&
(b)
\end{array}
$$
\caption{
$S(\rvx:\rvy|\rvlam^d)$
is nonzero 
for the qbnet of panel $(a)$
but zero for the qbnet of panel $(b)$.}
\label{fig-2-squares}
\end{figure}
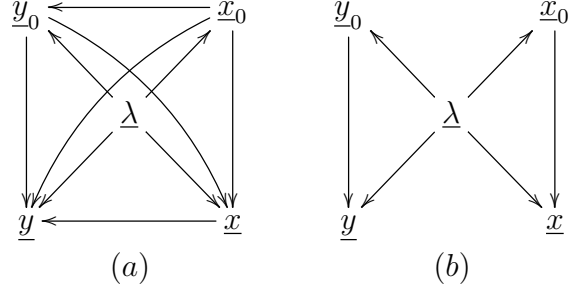
Eq.(\ref{eq-full-a-sq-ent})
can be represented
graphically
by the qbnet
 Fig.\ref{fig-2-squares}$(a)$.
The TPMs, printed in blue,
for the 2 qbnets $(a)$ and $(b)$ of
Fig.\ref{fig-2-squares},
are as follows.

\beq\color{blue}
A(\lam)=\sqrt{P(\lam)}
\eeq

\beq\color{blue}
A(x_0|\lam)=
\sqrt{P(x_0|\lam)}
\eeq

\beq\color{blue}
A(y_0|x_0, \lam)=
\left\{
\begin{array}{ll}
\sqrt{P(y_0|x_0,\lam)}
&\text{for Fig.\ref{fig-2-squares}$(a)$}
\\
\sqrt{P(y_0|\lam)}
&\text{for Fig.\ref{fig-2-squares}$(b)$}
\end{array}
\right.
\eeq

\beq\color{blue}
A(x|R_0, \lam)=
\left\{
\begin{array}{ll}
 A(x|R_0, \lam)
&
\text{for Fig.\ref{fig-2-squares}$(a)$}
\\
A(x|x_0, \lam)
&
\text{for Fig.\ref{fig-2-squares}$(b)$ }
\end{array}
\right.
\eeq

\beq\color{blue}
A(y|x, R_0, \lam)= 
\left\{
\begin{array}{ll}
A(y|x, R_0, \lam)
&
\text{for Fig.\ref{fig-2-squares}$(a)$}
\\
A(y|y_0, \lam)
&
\text{for Fig.\ref{fig-2-squares}$(b)$ }
\end{array}
\right.
\eeq

It's easy to check that
\begin{itemize}
\item
$\rvx\perp_G \rvy|\rvlam$
is false and
$S(\rvx:\rvy|\rvlam^d)\neq 0$
in Fig.\ref{fig-2-squares}$(a)$,
whereas
\item
$\rvx\perp_G \rvy|\rvlam$
is true and
$S(\rvx:\rvy|\rvlam^d)=0$
in Fig.\ref{fig-2-squares}$(b)$.
\end{itemize}

So far,
we have shown how, given 
any density  matrix
$\rho\in \cald_{\rvx, \rvy, \rvlam}$,
one can construct a
qbnet.
This method
of constructing a qbnet
from a density
matrix $\rho$ (or vice versa,
constructing a $\rho$
from a qbnet)
can be
generalized
to
finding
a qbnet
for any
$\rho\in\cald_{\rvx^{nx}}$.
for arbitrary $nx$.
Now we argue
that 
the
proof
of the quantum
d-separation
theorem
should be formally
identical 
to the
proof of the classical
d-separation  theorem.
The only
difference
between
the proofs is that
whenever a probability
occurs in
the classical proof, 
it must be replaced
by a vector amplitude
in the quantum proof.
Of course, probabilities
and vector amplitudes are
normalized
differently,
but that should
not change
the form of the proofs.
Note
that we have been
careful
to show
that vector amplitudes
can be conditioned and
satisfy a splitting rule, 
just like probabilities do.
Also,
we have been careful
to define 
d-separation 
$\rvA.\perp_G \rvB.|\rvZ.$
to be identical
for the classical and quantum cases.
Hence,
 we argue that, 
without
looking
at the details
of the
proof
of the classical
d-separation
theorem, 
one can conclude that
the following
theorem must be true:

\begin{framed}
\begin{claim}(Quantum d-separation Theorem)

Suppose
$\rvA., \rvB., \rvZ.$
are disjoint multinodes
of a DAG  $G$.

$\rvA.\perp_G \rvB.|\rvZ.$ iff
$S_\rho(\rvA. : \rvB.|\rvZ.^d)=0$
for all $\rho$
compatible with $G$.

\end{claim}
\end{framed}

Define the squashed
entanglement
of a density
matrix $\rho_{\rvx, \rvy}$ by

\beq
E_{sq}(\rho_{\rvx,\rvy})=
\frac{1}{2}
\min_{\rho\in \cald}
S_\rho(\rvx:\rvy|\rvlam^d)
\eeq
where
$\cald=\{\rho
\in
\cald_{\rvx, \rvy, \rvlam^d}\;\;|\;\;
\tr_{\rvlam^d}
\rho_{\rvx,\rvy, \rvlam^d}
=
\rho_{\rvx, \rvy}
\}$.
Then the quantum 
d-separation
theorem immediately(?) implies
the following.

\begin{claim}
Suppose
$\rvA., \rvB.$
are disjoint multinodes
of a DAG  $G$,
 
($\rvA.\perp_G \rvB.|\rvZ.$ 
for some $\rvZ.$ such that
$\rvA., \rvB., \rvZ.$
are disjoint multinodes of G)
iff
$E_{sq}(\rho_{\rvA., \rvB.})=0$.
\end{claim}

\section{Quantum Belief Propagation}

The rules
(and their proof)
of classical BP
can be found
in the Bayesuvius chapter
entitled ``Message
Passing (Belief Propagation)".
Just like
the proof
of the classical 
d-separation
theorem, 
the
proof
of the rules for classical BP
 relies on 3
ingredients:
\begin{itemize}
\item the topology
of a
DAG
\item
the definition
of conditional
probabilities 
as ratios of 
joint
probabilities
\item
the splitting rule
for probabilities
\end{itemize}
Since these
3 ingredients
are also
available
in the quantum
side
if we replace
probabilities
by vector amplitudes,
we can
conclude that
the rules
for quantum
BP
are formally 
the same as those
for classical BP,
modulus the
replacement of
probabilities
by
vector amplitudes.
The
difference in
normalization
of probabilities
and 
vector amplitudes
does not
make
the 
rules
for classical
and quantum
BP
different
because
these
rules
are defined up
to a normalization
constant.

In Bayesuvius, 
the BP chapter entitled ``Message Passing (Belief Propagation)"
considers a general case of classical BP (viz., BP for
polytrees (BP-Gen)) and a special case of classical
BP (viz., BP for bipartite bnets (BP-BB)).
We end this section
with 2 subsections dedicated
to the quantum
analogues of the rules
for BP-Gen and BP-BB.
Those 2 subsections
are simply exact quotes
from the BP chapter in
Bayesuvius,
except that
all $P$'s have been
crossed out and replaced by $\A$'s.

\subsection{Quantum BP for polytrees}
Let $\rva^{na}=
(\rva_i)_{i=0, 1, \ldots, na-1}$
denote the parents of $\rvx$
and
$\rvb^{nb}=
(\rvb_i)_{i=0, 1, \ldots, nb-1}$
its children.

Define

\beqa
\label{eq-mp-pix}
\pi_\rvx(x)&=&
\sum_{a^{na}}\AP(x|a^{na})
\prod_i
\pi_{\rvx\ldart\rva_i}
(a_i)\\
&=&E_{\rva^{na}}[\AP(x|a^{na})]
\eeqa
(boundary case: if $\rvx$
is a root node, use $\pi_\rvx(x)=\AP(x)$.)
and

\beq
\lam_\rvx(x)=
\prod_i
\lam_{\rvb_i\rdart \rvx}(x)
\;.
\label{eq-mp-lamx}
\eeq
(boundary case: if $\rvx$
is a leaf node, use $\lam_\rvx(x)=1$.)

\begin{itemize}

\item{\bf RULE 1: (red parent)}

\beqa
\label{eq-mp-rule1}
\underbrace{\lam_{\rvx\rdart\rva_i}
(a_i)}_{OUT}&=&
\caln(!a_i)
\sum_x\left[
\underbrace{\lam_\rvx(x)}_{IN}
\sum_{(a_k)_{k\neq i}}\left(
\AP(x|a^{na})\prod_{k\neq i}
\underbrace{\pi_
{\rvx\ldart\rva_k}
(a_k)}_{IN}
\right)\right]
\\&=&
\caln(!a_i)
\sum_x\left[
\lam_\rvx(x)
E_{(\rva_k)_{k\neq i}}[\AP(x|a^{na})]\right]
\\&=&
\caln(!a_i)
E_{(\rva_k)_{k\neq i}}E_{\rvx|a^{na}}
\lam_\rvx(x)
\eeqa
(boundary case:
if $\rvx$ is a root node, use
$\lam_{\rvx\rdart\rva_i}
(a_i)=\caln(!a_i)$.)

\item{\bf RULE 2: (red child)}

\beq
\underbrace{\pi_{\rvb_i\ldart\rvx}
(x)}_{OUT}=
\caln(!x)
\underbrace{\pi_\rvx(x)}_{IN}
\prod_{k\neq i}
\underbrace{\lam_{\rvb_k\rdart \rvx}(x)}_{IN}
\label{eq-mp-rule2}
\eeq
(boundary case:
if $\rvx$ is a leaf node, use
$\pi_{\rvb_i\ldart\rvx}(x)=
\caln(!x)\pi_\rvx(x)$
.)

\end{itemize}
In the above
equations, if the
range set of a product is empty, then
 define the product as 1; i.e.,
$\prod_{k\in \emptyset}F(k)=1$.

\hrule\noindent
{\bf Claim:} Define

\beq
BEL^{(t)}(x)=\caln(!x)
\lam^{(t)}_\rvx(x)
\pi^{(t)}_\rvx(x)
\;.\eeq
Then

\beq
\lim_{t\rarrow \infty}
BEL^{(t)}(x)=\AP(x|\eps)
\;.
\eeq
This  says that
the belief in $\rvx=x$
converges to $\AP(x|\eps)$ and it
equals the product
of messages received from all
parents and children of $\rvx=x$.
\subsection{Quantum BP for Bipartite Bnets}

\begin{enumerate}

\item {\bf
Traversing an $x$ (i.e., root) node.}

For $i=0, 1, \ldots , nx-1$, if
 $\alp\in nb(i)$, then,

\beq
m^{(t)}_{\alp\ldart i }(x_i)
=
\prod_{
\beta\in nb(i)-\alpha}
m^{(t-1)}_{\beta\rdart i}(x_i)
\;,
\label{eq-mp-iter1}
\eeq
whereas if  $\alp\notin nb(i)$

\beq
m^{(t)}_{\alp\ldart i}(x_i)=
m^{(t-1)}_{\alp\ldart i}(x_i)
\;.
\eeq

\item {\bf
Traversing an $f$ (i.e., leaf) node.}

For $\alp=0, 1, \ldots, nf-1$, if
 $i\in nb(\alp)$, then

\beqa
m^{(t)}_{\alp\rdart i}(x_i)
&=&
\sum_{(x_k)_{k\in nb(\alpha)-i}}
f_\alpha(x_{nb(\alpha)})
\prod_{k\in nb(\alpha)-i}
m^{(t-1)}_{\alp\ldart k }
(x_k)
\\
&=&
E^{(t-1)}_{(x_k)_{k\in nb(\alpha)-i}}[
f_\alpha(x_{nb(\alpha)})]
\;,
\label{eq-mp-iter2}
\eeqa
whereas if $i\notin nb(\alp)$

\beq
m^{(t)}_{\alp\rdart i}(x_i)
=
m^{(t-1)}_{\alp\rdart i}(x_i)
\;.
\eeq

\end{enumerate}

In the above
equations, if the
range set of a product is empty, then
 define the product as 1; i.e.,
$\prod_{k\in \emptyset}F(k)=1$.

\hrule\noindent
{\bf Claim:}

\beq
\AP(x_i|\eps)=
\lim_{t\rarrow
\infty}\caln(!x_i)\prod_{\alp\in nb(i)}
m^{(t)}_{\alp\rdart i}(x_i)
\;
\label{eq-m-prod}
\eeq
and

\beq
\AP(x_{nb(\alp)}|\eps)=\lim_{t\rarrow \infty}
\caln(!x_{nb(\alp)})
f_\alp(x_{nb(\alp)})
\prod_{k\in nb(\alp)}
m^{(t)}_{\alp\ldart k}(x_k)
\;.
\label{eq-f-m-prod}
\eeq
\hrule
{\color{red}In the classical
case, 
$f_\alp(x_{nb(\alp)})$
stands for a real valued
function (e.g., $P(f_\alp=1|x_{nb(\alp)})$),
whereas in the quantum case,
it stands for a vector amplitude
(e.g.,  $\A(f_\alp=1|x_{nb(\alp)})$).}

\appendix
\section{Appendix: Reduced qbnet}
\begin{figure}[h!]
$$
\begin{array}{ccc}
\xymatrix{
&\rvlam\ar[dl]\ar[dr]
\\
\rvY&&\rvX\ar[ll]
}
&\;\;\;&
\xymatrix{
&\rvlam\ar[dl]\ar[dr]
\\
\rvY&&\rvX
}
\\
(a)
&&
(b)
\end{array}
$$
\caption{The 2 qbnets in
Fig.\ref{fig-2-squares}
can sometimes be reduced to these 2 qbnets.}
\label{fig-2-triangles}
\end{figure}
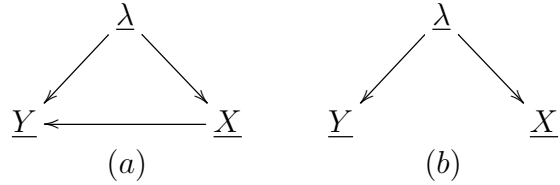

Let $\rvX=(\rvx, \rvx_0)$ and 
$\rvY=(\rvy, \rvy_0)$.
Note that 
if, as
indicated in
Eq.(\ref{eq-y0-indep}),
$A(x|R_0,\lam)$ is
independent of $y_0$, then  
the 2 qbnets in  Fig.\ref{fig-2-squares}
can be reduced to the 2 qbnets in Fig.\ref{fig-2-triangles},
The node TPMs, printed in blue,
of the qbnets in Fig.\ref{fig-2-triangles},
are as follows:

\beq\color{blue}
A(\lam)=\sqrt{P(\lam)}
\eeq

\beq\color{blue}
A(x,x_0|{\color{red}\cancel{y_0}}
, \lam)=
\left\{
\begin{array}{ll}
 A(x|{\color{red}\cancelto {x_0}{R_0}}
, \lam)\sqrt{P(x_0|\lam)}
&
\text{for Fig.\ref{fig-2-triangles}$(a)$}
\\
A(x|x_0, \lam)\sqrt{P(x_0|\lam)}
&
\text{for Fig.\ref{fig-2-triangles}$(b)$ }
\end{array}
\right.
\label{eq-y0-indep}
\eeq

\beq\color{blue}
A(y, y_0|x, x_0, \lam)= 
\left\{
\begin{array}{ll}
A(y|x, R_0, \lam)\sqrt{P(y_0|x_0,\lam)}
&
\text{for Fig.\ref{fig-2-triangles}$(a)$}
\\
A(y|y_0, \lam)\sqrt{P(y_0|\lam)}
&
\text{for Fig.\ref{fig-2-triangles}$(b)$ }
\end{array}
\right.
\eeq

\bibliographystyle{plain}
\bibliography{references}
\end{document}